\documentclass[aps,prl,twocolumn,superscriptaddress]{revtex4}
\usepackage{mathrsfs}
\usepackage{epsfig}
\usepackage{graphicx}
\usepackage{amsfonts}
\usepackage[figuresright]{rotating}
\usepackage{amssymb}
\usepackage{amsmath}
\usepackage{dcolumn}
\usepackage{bm}
\usepackage{color}

\def\be{\begin{equation}} \def\ee{\end{equation}}
\def\bea{\begin{eqnarray}} \def\eea{\end{eqnarray}}

\newcommand{\WQCASQC} {Wilczek Quantum Center and Key Laboratory of Artificial Structures and Quantum Control, School of Physics and Astronomy, Shanghai Jiao Tong University, Shanghai 200240, China}

\newcommand{\SRCQC}{Shanghai Research Center for Quantum Sciences, Shanghai 201315, China}

\begin{document}
%\title{Imaginary time crystal: {\clb an exotic phase of quantum matter}}
\title{A $\frac 13$ power-law universality class out of  stochastic driving in interacting systems}

\author{Zi Cai}
\email{zcai@sjtu.edu.cn}
\affiliation{\WQCASQC}
\affiliation{\SRCQC}

\begin{abstract}  % Science journal - 125 word or less; current 119.

In this paper, we study the mean-field dynamics of a general class of many-body systems with stochastically fluctuating interactions. Our findings reveal a universal algebraic decay of the order parameter $m(t)\sim t^{-\chi}$ with an exponent $\chi=\frac 13$ that is independent of most system details including the strength of the stochastic driving, the energy spectrum of the undriven systems,  the initial states and even the driving protocols. It is shown that such a dynamical universality class can be understood as a consequence of a diffusive process with a time-dependent diffusion coefficient which is  determined self-consistently during the evolution.  The finite-size effect, as well as the relevance of our results with current experiments in high-finesse cavity QED systems are also discussed.

\end{abstract}

%\pacs{05.30.Jp, 75.10.Pq, 02.70.Ss, 03.65. Yz}

\maketitle

{\it Introduction --}
A universality class is a collection of diverse systems which share common properties. For example, systems with dramatically different microstructures may share  identical critical exponents near the phase transition point, where not only the static properties\cite{Wilson1983} but also the relaxation dynamics\cite{Hohenberg1977} might exhibit similar asymptotic behavior that is independent of most  system details.  Compared to equilibrium systems, the universality class in systems operating far from equilibrium is significantly richer but less known in general. As a prototypical example,  the Kardar-Parisi-Zhang (KPZ) universality class\cite{Kardar1986} with a dynamic critical exponent $z=\frac 32$ governs diverse non-equilibrium phenomena from surface growth  in classical stochastic models\cite{Halpin1995} to super-diffusion in integrable quantum models\cite{Ljubotina2019,Nardis2021,Wei2021}.

Random fluctuations, almost by definition, are usually considered  a source of disorder that leads to irregular spatiotemporal behavior.  A profound question is under the random fluctuation, do systems always adjust their macroscopic behavior to the average properties of the fluctuation, or can one find certain systems responding to randomness in a more active way, thus exhibiting, for instance, nontrivial behavior which is forbidden in the corresponding deterministic systems without randomness.  The answer is indeed positive. It has been shown that random fluctuation in certain nonlinear models  can conspire with nonlinearity to render counterintuitive behavior such as noise-induced spatiotemporal order and phase transition\cite{Sagues2007}. The key point here involves the randomness that is introduced externally through a stochastic process of a control parameter that gets multiplied into  system variables, and thus is multiplicative instead of additive as  in the common Langevin treatment of the internal fluctuation. Such noise-enhanced regularity phenomena have been extensively studied in  nonlinear systems ranging from electronic to biochemical systems\cite{Horsthemke1984}. A more profound question, then, is whether such a   multiplicative noise can give rise to nontrivial dynamical universality due to the conspiracy of stochasticity and nonlinearity.

In this paper, we attempt to answer this question by focusing for simplicity on the mean-field dynamics of a general class of interacting systems with stochastic driving. Generally, a system can be driven out of equilibrium via a time-dependent manipulation of the Hamiltonian parameters. The majority of studies in this field has focused on  the cases where the Hamiltonian parameters are tuned in a regular way, for instance, they can be  suddenly\cite{Mitra2018,Alessio2015}, linearly\cite{Kibble1980,Campo2014} or periodically\cite{Goldman2014,Eckardt2017,Harper2020,Oka2019} changed in time, corresponding to  the quench, ramping or periodically-driven dynamics respectively. However, only a few efforts studies have been focusing on stochastically-driven systems\cite{Marino2012,Halimeh2018,Nandy2017,Nandy2018,Kos2021}. Perhaps this lack of research studies on this field can be justified by the fact that  such  randomness in the time domain will heat the system, and thus it is bound to be detrimental to any spatiotemporal order\cite{Hu2015,Buchhold2015,Cai2017}. However, motivated by the multiplicative noise in the classical systems, this paper reveals that a stochastic driving acting on the interaction strength instead of external field,  may facilitate an intriguing dynamical behavior, where the order parameter of the  system exhibits a universal algebraic decay in time $t^{-\chi}$ with an exponent $\chi=\frac 13$ that is independent of the noise strength, the energy spectrum of the undriven systems, the initial states and even the driving protocols. It is shown that such universal dynamics can be understood as consequence of a nonlinear diffusion process with a time-dependent diffusion coefficient that is determined self-consistently during the time evolution.

\begin{figure}[htb]
\includegraphics[width=0.99\linewidth]{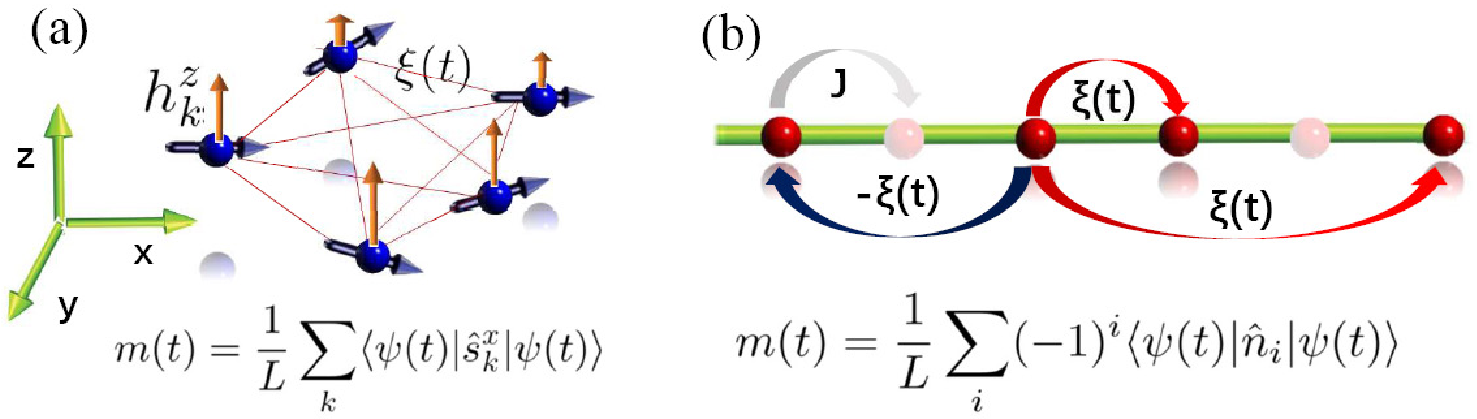}
\includegraphics[width=0.99\linewidth,bb=206 27 1290 346]{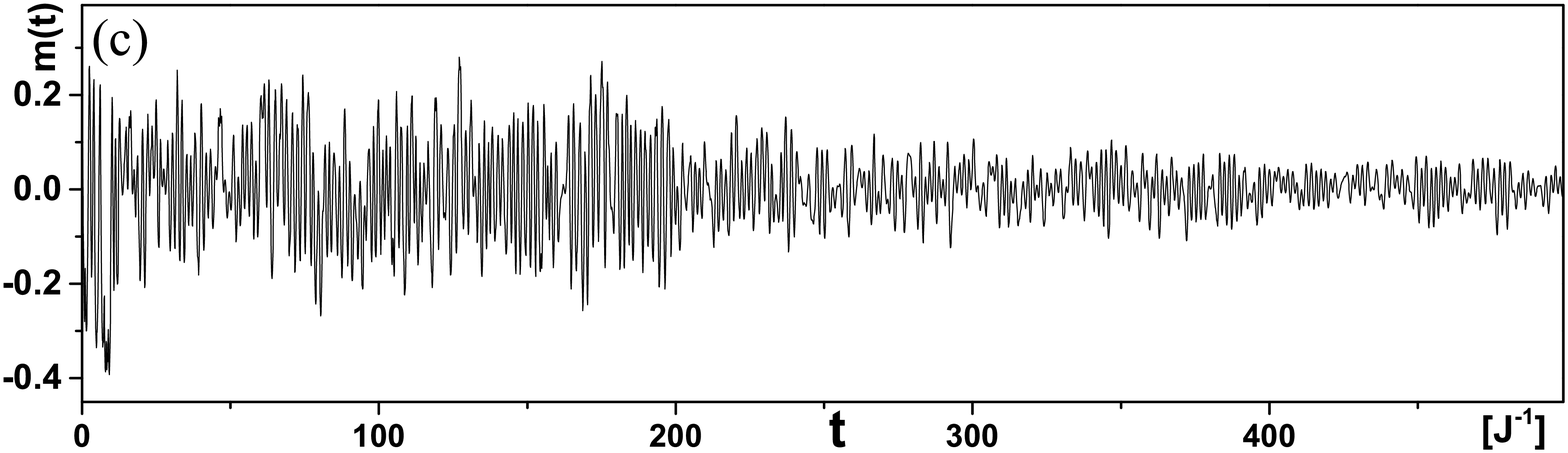}
\caption{Schematic diagram of (a) a fully-connected transverse Ising model and (b) a 1D hard-core bosonic model with infinite-range interactions;
(c) the dynamics of the order parameter for a single random trajectory with L=5000, $\mathcal{D}=5J$.}
\label{fig:fig1}
\end{figure}

{\it Model I --} We first consider a fully-connected transverse Ising model\cite{Titum2020} subjected to a non-uniform magnetic field (Fig.\ref{fig:fig1} a), whose Hamiltonian reads:
\begin{equation}
H_s(t)=-\frac{\xi(t)}L\sum_{kk'} \hat{s}_k^x\hat{s}_{k'}^x-\sum_k h^z_k \hat{s}_k^z \label{eq:Hamspin}
\end{equation}
where $\hat{s}^\alpha_k$  is the spin-$\frac 12$ operator operating on $k$th site(mode) defined as $\hat{s}^\alpha_k=\frac 12\hat{\sigma}^\alpha_k$ ($\hat{\sigma}^\alpha_k$ are the three Pauli matrices with $\alpha=x,y,z$). $h^z_k$ is the magnetic field along z-direction on  site(mode) $k$ which will be specified later on. $L$ is the total number of spins  and the $\frac 1L$ pre-factor in Eq.(\ref{eq:Hamspin}) guarantees that the total interacting energy scale linearly with $L$.  $\xi(t)$ is the strength of the all-to-all interaction that is spatially unform but is randomly fluctuating in time, satisfying: $\langle \xi(t)\rangle_{\bm\xi}=0$, $\langle \xi(t)\xi(t')\rangle_{\bm\xi}=\mathcal{D}^2 \delta(t-t')$ with $\mathcal{D}$ being the strength of the stochastic driving. The ensemble average $\langle\cdot\rangle_{\bm\xi}$ is over all random trajectories of the stochastic driving.

Due to this all-to-all feature, the interaction terms in Eq.(\ref{eq:Hamspin}) can be decoupled by introducing a time-dependent ferromagnetic order parameters
\begin{equation}
m(t)=\frac 1L\sum_k\langle\psi(t)|\hat{s}_k^x|\psi(t)\rangle \label{eq:OP}
%m(t)=\frac 1L\sum_i (-1)^i \langle\psi(t)|\hat{n}_i|\psi(t)\rangle \label{eq:OP}
\end{equation}
with $|\psi(t)\rangle$ representing the wavefunction of the system at time t. The mean-field(MF) Hamiltonian turns to a set of spin-$\frac 12$ systems as $\tilde{H}_s(t)=\sum_k \tilde{H}_k(t)$ with :
\begin{small}
\begin{equation}
\tilde{H}_k(t)=-\xi(t)m(t)\hat{s}_k^x- h^z_k \hat{s}_k^z\label{eq:MFspin},
\end{equation}
\end{small}
and the corresponding equation of motion(EOM) reads:
\begin{equation}
\dot{\mathbf{s}}_k=\mathbf{h}_k(t)\times \mathbf{s}_k\label{eq:EOM}
\end{equation}
where $\mathbf{s}_k=[\langle\hat{s}_k^x\rangle,\langle\hat{s}_k^y\rangle,\langle\hat{s}_k^z\rangle]^T$ is a vector in the Bloch sphere ($|\mathbf{s}_k|=\frac 12$).  $\mathbf{h}_k=[-\xi(t)m(t),0,h_k^z]$ and $m(t)$ is self-consistently determined by Eq.(\ref{eq:OP})  during  evolution. Here, the all-to-all couplings in Eq.(\ref{eq:Hamspin}) guarantee that spatial fluctuation in the thermodynamical limit ($L\rightarrow \infty$)  is completely suppressed, and thus  the MF Hamiltonian.(\ref{eq:MFspin}) and EOM.(\ref{eq:EOM})are no longer approximations but the exact methods dealing with both the equilibrium and  non-equilibrium systems\cite{Blass2018,Igloi2018,Supplementary}. Notice that in the MF treatment, the EOM(\ref{eq:EOM}) does not only apply to the quantum cases, but also to the classical cases, where the $s_k$ in Eq.(\ref{eq:Hamspin})is a classical vector instead of operator.  In general, the MF Hamiltonian.(\ref{eq:MFspin}) and the EOM.(\ref{eq:EOM})  can  not only be applied to quantum or classical magnetism (Eq.(\ref{eq:Hamspin}) for instance), but also to diverse phenomena with spontaneous symmetry breaking ranging from superconductors\cite{Anderson1958,Barankov2004,Yuzbashyan2005,Barankov2006,Yuzbashyan2006} to charge-density-waves (CDW)\cite{Chen2020}.

{\it Model II --} In most realistic systems with local interactions, the spatial fluctuations render the mean-field EOM.(\ref{eq:EOM}) as no longer exact. Despite this fact, recent experimental progress in high-finesse cavity QED systems have opened up new possibilities to study infinite-range interactions\cite{Landig2016,Hruby2018}. For instance, it is possible to realize a one-dimensional(1D) hard-core bosonic system with an infinite-range interaction (Fig.\ref{fig:fig1} b) with Hamiltonian:
\begin{small}
\begin{equation}
H_b=-J\sum_i (\hat{b}_i^\dag \hat{b}_{i+1}+h.c)-\frac{\xi(t)}L \sum_{ij}(-1)^{i-j} \hat{n}_i \hat{n}_j \label{eq:Hamboson}
\end{equation}
\end{small}
where $J$ is the nearest-neighbor hopping amplitude, $b_i^\dag$ ($b_i$) is the creation (annihilation) operators of the hard-core boson at site i, and $\hat{n}_i=\hat{b}_i^\dag \hat{b}_i$. The interaction $\xi(t)$ is defined the same as in Eq.(\ref{eq:Hamspin}). If $\xi(t)> 0$, the interaction in Eq.(\ref{eq:Hamboson}) is attractive(repulsive) between two bosons in the same(different) sublattice, and thus favors a CDW state at half-filling. As a consequence, one can introduce a CDW order parameter $m(t)=\frac 1L \sum_i (-1)^i \langle\hat{n}_i\rangle$ to decouple the interaction, which yields to a MF Hamiltonian: $\tilde{H}_b=-J\sum_i (\hat{b}_i^\dag \hat{b}_{i+1}+h.c) +m(t)\xi(t)\sum_i (-1)^i \hat{n}_i$. One can further use the Jordan-Wigner transformation to transform the 1D hard-core bosons to spinless fermions: $\hat{b}_i^\dag= e^{i\sum_{j<i}\pi \hat{n}_j}\hat{c}_i^\dag$, then perform the Fourier transformation  $c^\dag_i=\frac{1}{\sqrt{L}}\sum_k e^{ik i} c^\dag_k$ to transfer the MF Hamiltonian into the momentum space as $\tilde{H}_b(t)=\sum_k \Psi_k^\dag \hat{H}_k(t) \Psi_k$ where $\Psi_k=[c_k,c_{k+\pi}]^T$ and $\hat{H}_k(t)$ has exactly the same form of Eq.(\ref{eq:MFspin}) provided that we assume $h_k^z=-2J\cos k$ with $k\in [-\pi,\pi]$. Therefore, the mean-field EOM.(\ref{eq:EOM}) also provides an exact description of the dynamics of this interacting bosonic system that is closely related with the current cavity QED experiments. In the following, we choose $h_k^z=-2J\cos k$ unless it is specified otherwise.

Numerically, we adopt Stratonovich's formula\cite{Kloeden1995} of the stochastic differential Eq.(\ref{eq:EOM}), and solve it using the standard Heun  method\cite{Ament2016} with the time step of $\Delta t=10^{-5}J^{-1}$, the convergence of which has been  numerically assessed\cite{Supplementary}.  The ensemble average over the random trajectories is performed by directly sampling over the $\mathcal{N}$ sets of noise realizations ($\mathcal{N}=500$ in our simulations).
\begin{figure*}[htb]
\includegraphics[width=0.325\linewidth,bb=72 53 713 525]{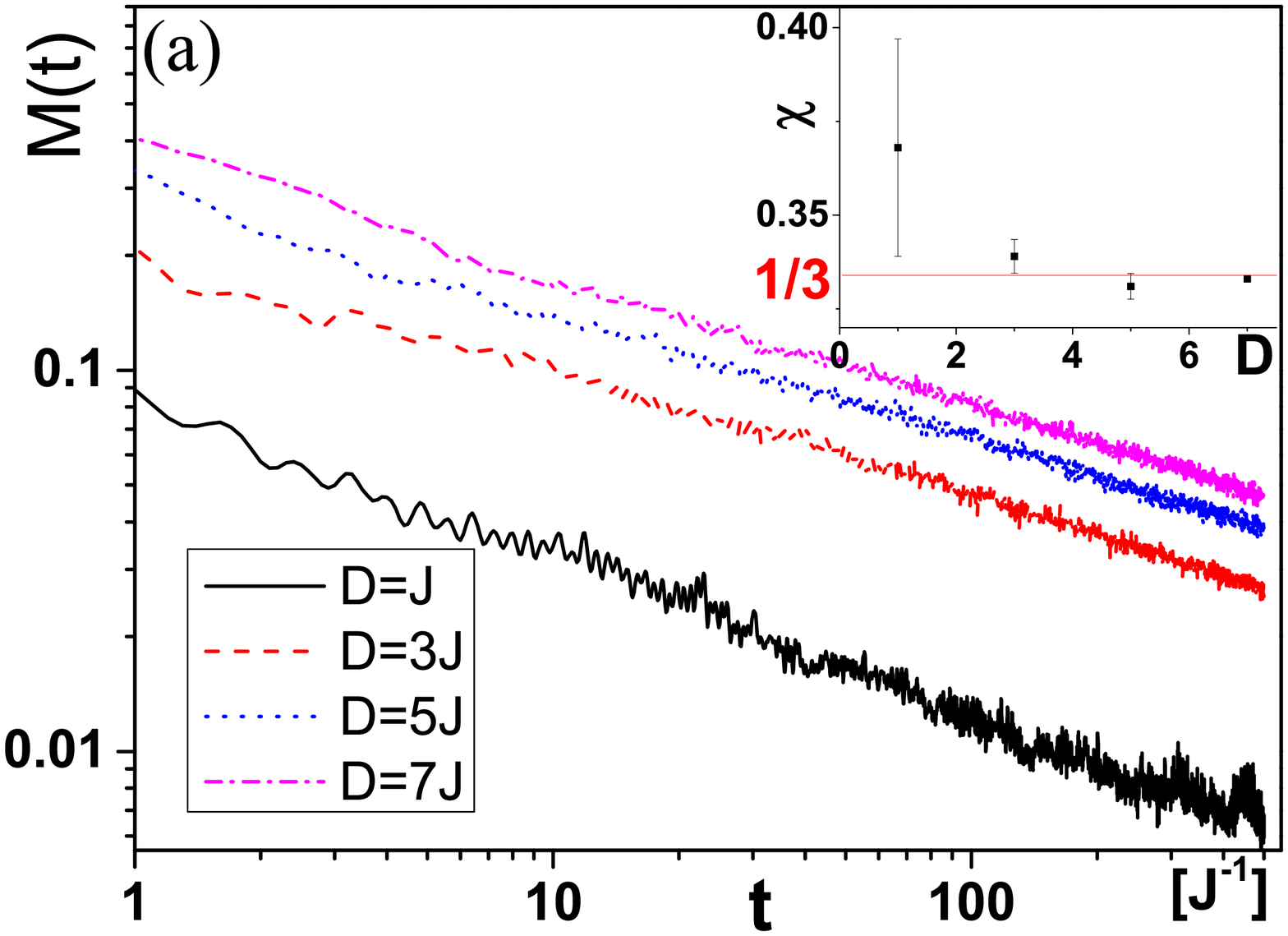}
\includegraphics[width=0.325\linewidth,bb=72 53 713 525]{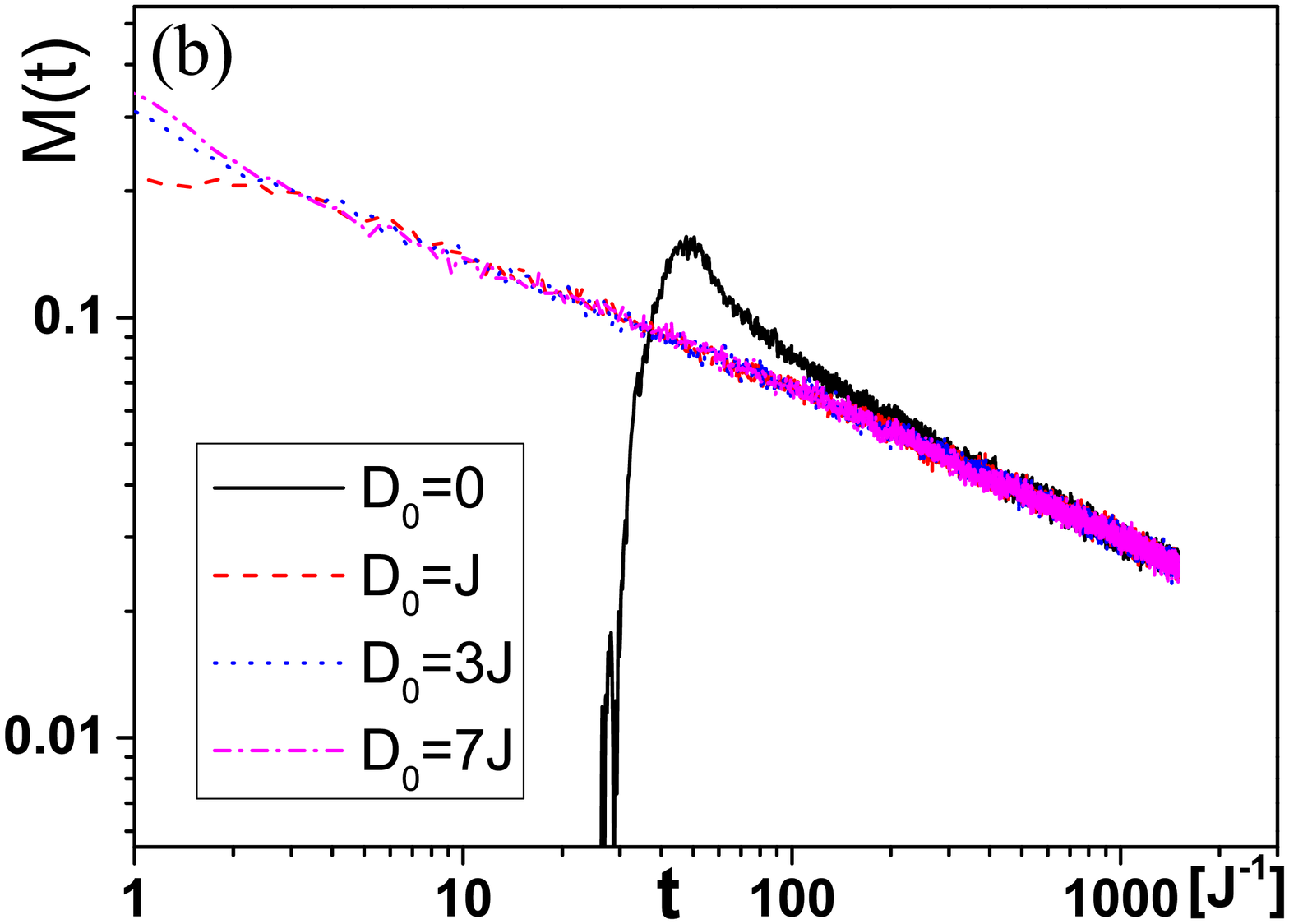}
\includegraphics[width=0.325\linewidth,bb=72 53 713 525]{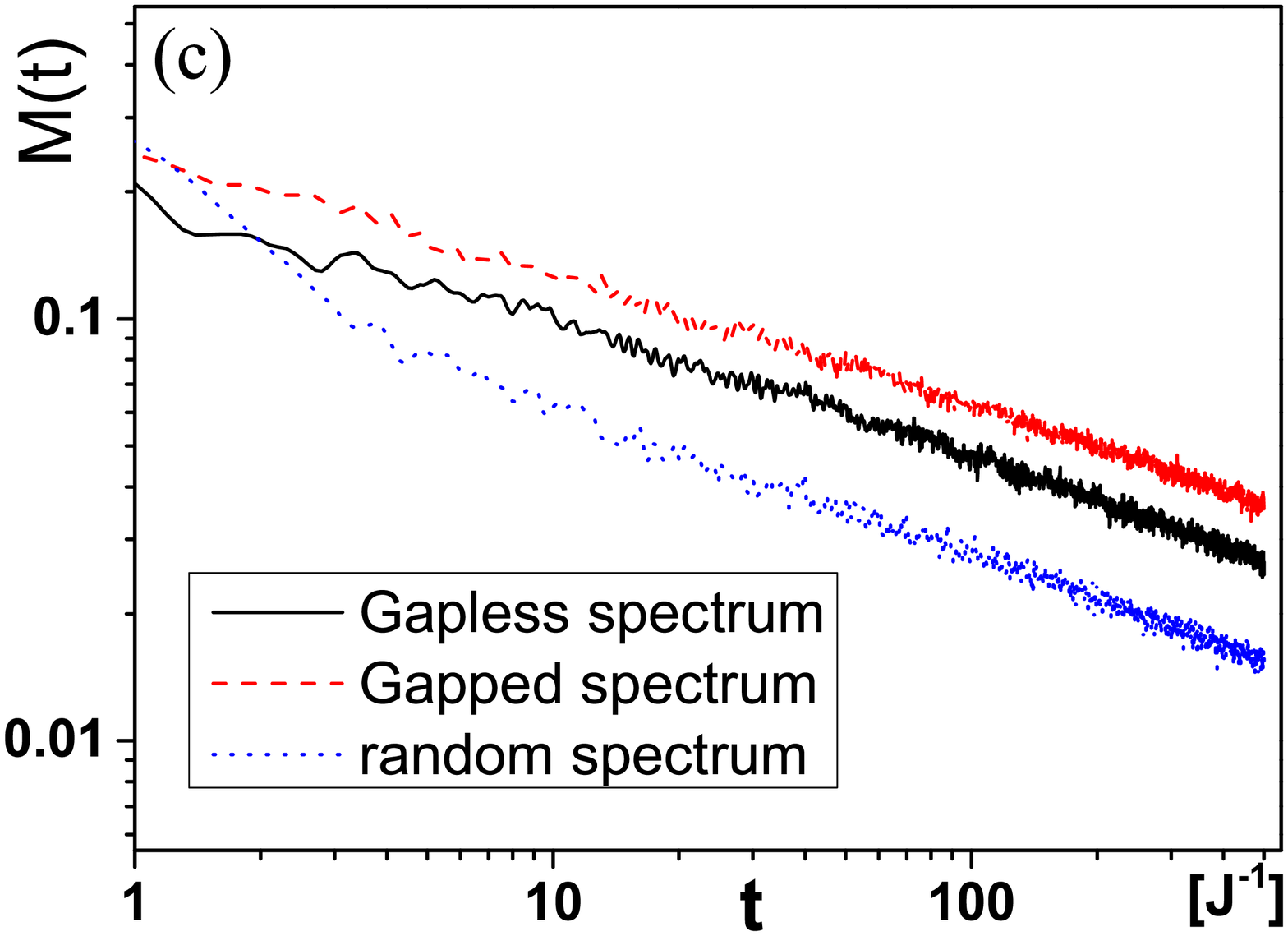}
\includegraphics[width=0.325\linewidth,bb=72 53 713 525]{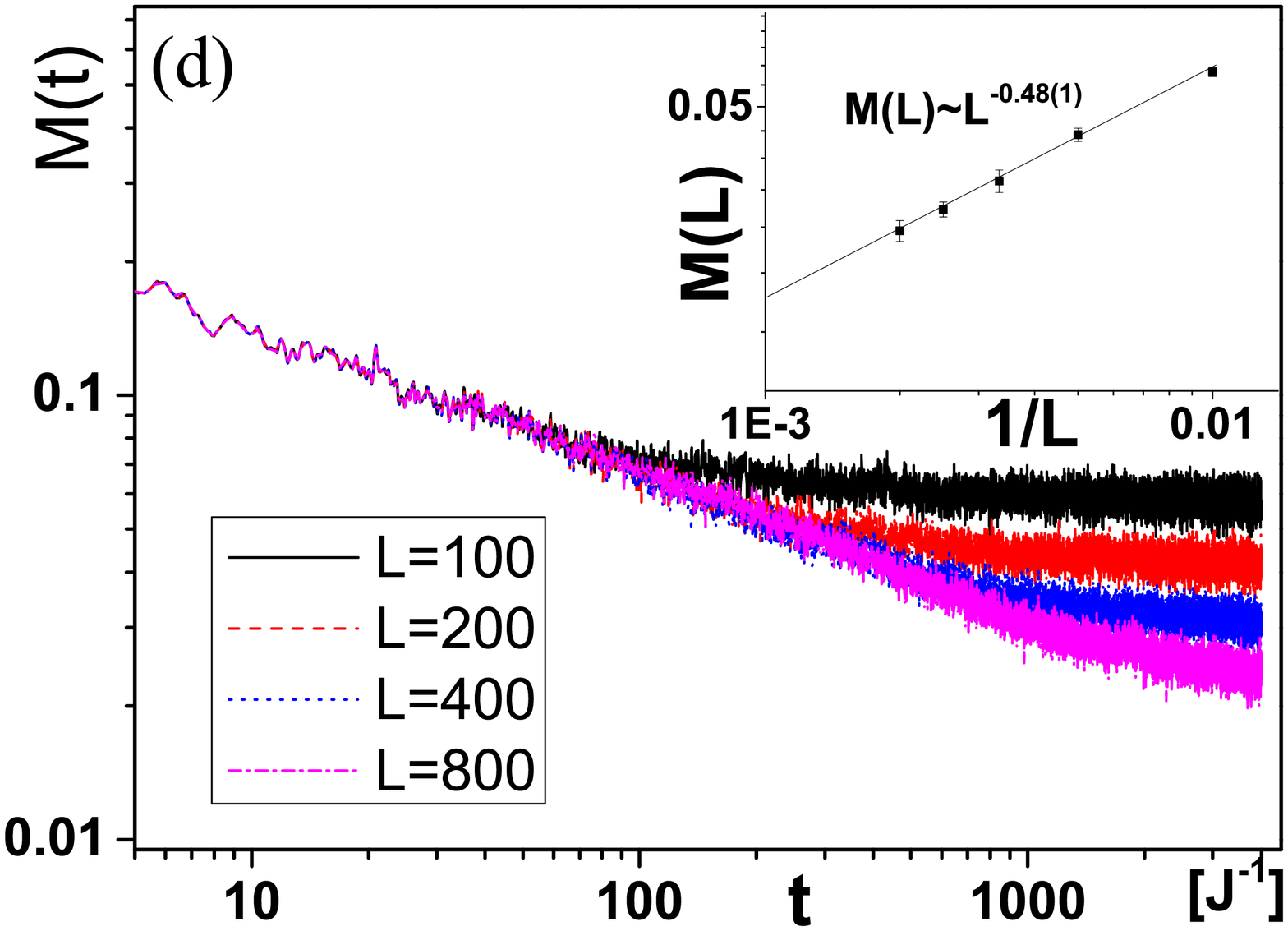}
\includegraphics[width=0.325\linewidth,bb=72 53 713 525]{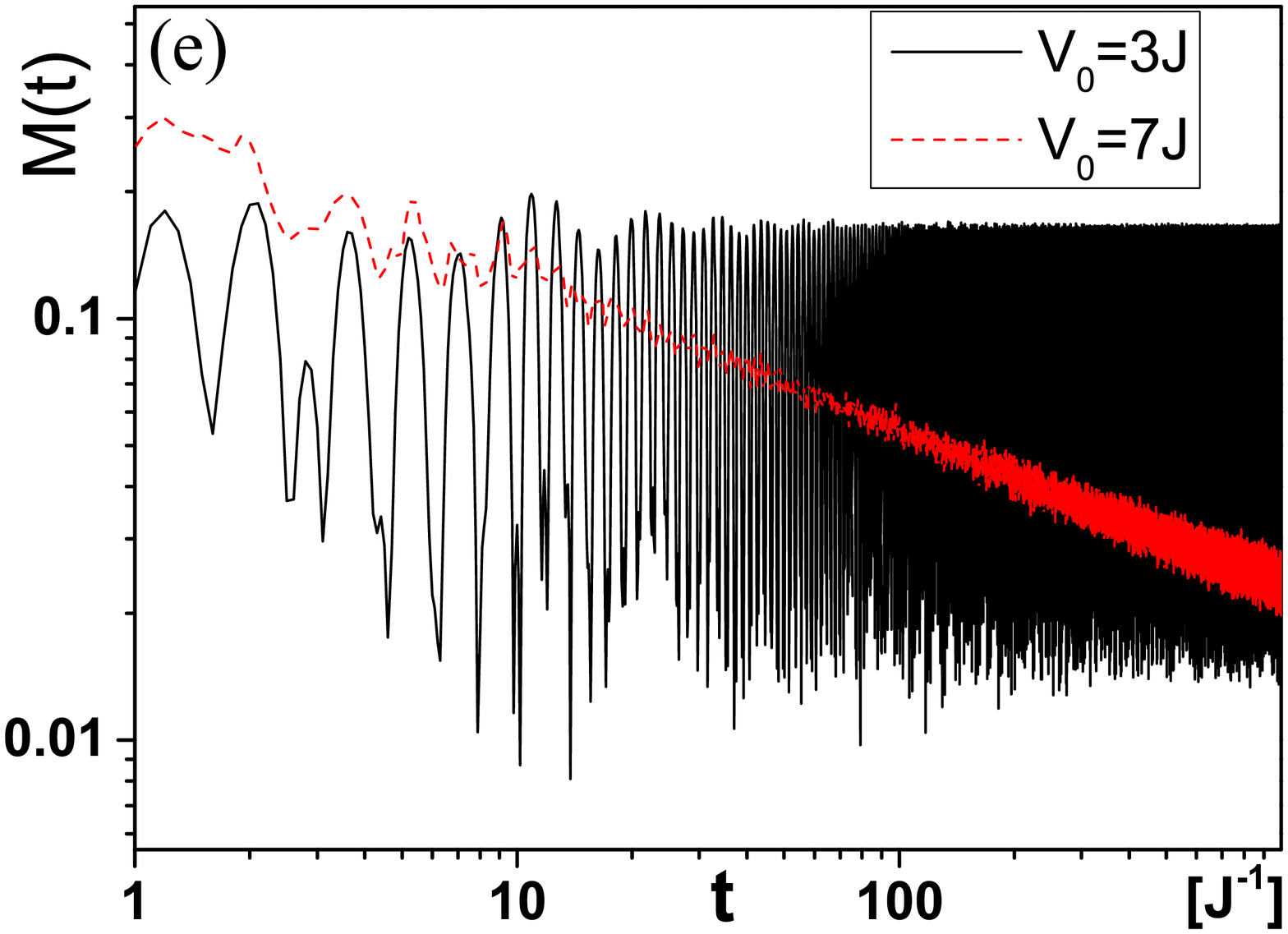}
\includegraphics[width=0.335\linewidth,bb=72 53 713 525]{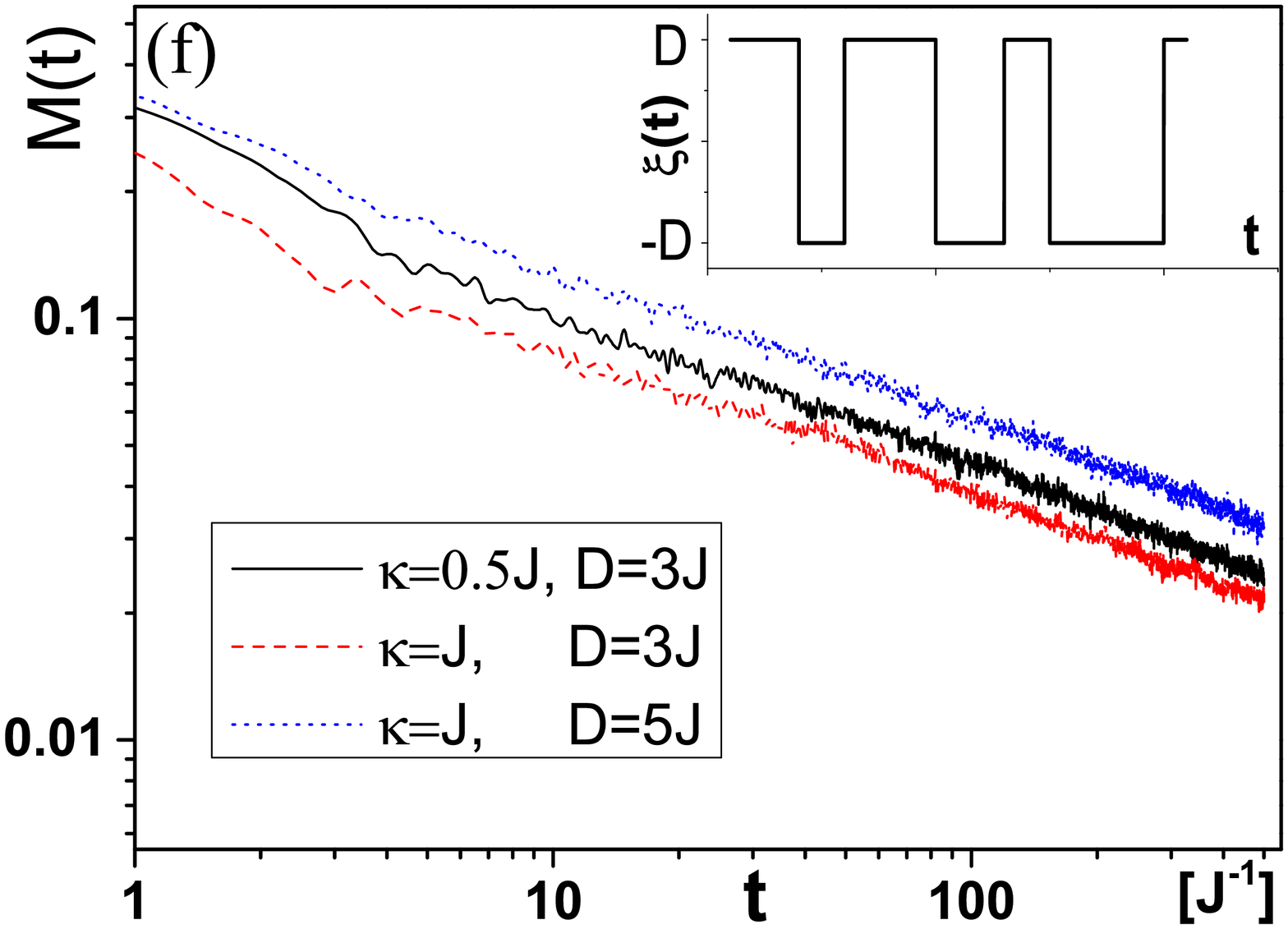}
\caption{(a) Dynamics of the order parameter amplitude $M(t)$ for systems with various stochastic driving strengths $\mathcal{D}$ (the inset represents the dependence of the power-law exponent $\chi$ on $\mathcal{D}$); (b)$M(t)$ starting from different initial states as the ground state of Hamiltonian.(\ref{eq:Hamspin}) with $\xi(t=0)=\mathcal{D}_0$; (c)$M(t)$ for system with gapless ($h_k^z=-2J\cos k$), gapped ($h_k^z=\pm\sqrt{(2J\cos k)^2+\Delta^2}$ with $\Delta=J$) and completely random ($h_k^z\in [-2J,2J]$)  spectrum functions; (d)$M(t)$ for finite systems (the inset is the saturated value $M(L)=M(t\rightarrow \infty)$ as a function of the system size $L$); (e)$M(t)$ in the presence of a quasi-periodic driving $\xi(t)=V_0[\cos(2\pi t+\varphi)+\cos(\sqrt{3} t)]$; (f)$M(t)$ in the presence of a telegraph stochastic driving with $\xi(t)$ randomly jumping between two discrete values $\mathcal{D}$ and $-\mathcal{D}$ with a transition rate $\kappa$ (the inset is a sketch of a single trajectory of the telegraph driving).      We choose $\Delta t=10^{-5}J^{-1}$; the systems size $L=5000$ except (d); the initial states are selected as the ground states of the Hamiltonian.(\ref{eq:Hamspin}) with $\xi(t=0)=\mathcal{D}$ except (b); $\mathcal{D}=3J$ for (c) and $\mathcal{D}=5J$ for (b) and (d).}
\label{fig:fig2}
\end{figure*}

{\it Universal algebraic decay of the amplitude to order parameter--} In this simulation, we focus on the dynamics of the order parameter $m(t)$.  In general, the stochastic driving will heat the system toward a featureless state with $m(t\rightarrow \infty)=0$, thus  for a single random trajectory  $\{\xi(t)\}$, m(t) will exhibit a chaotic oscillation whose amplitude decays in time, as shown in Fig.\ref{fig:fig1} (c).  Despite the triviality of the steady state, the asymptotic behavior approaching it can be highly non-trivial and exhibit intriguing universal behavior even at a mean-field level. To characterize such an universal dynamics, we calculate  the average amplitude of the order parameter at time t defined as $M(t)=\sqrt{\langle m^2(t)\rangle_{\bm\xi}}$, and we end up with a universal algebraic decay $M(t)\sim t^{-\frac 13}$ that is independent of most system details as we will show in the following.

We first check the dependence of the long-time behavior of $M(t)$ on the strength of the stochastic driving $\mathcal{D}$. As shown in Fig.\ref{fig:fig2} (a), during the time evolution $M(t)$ in the cases with strong stochastic driving are typically larger than those in weakly driven cases, which seems to indicate that this randomness facilitates the order phase. Contrary to many other studies, such a counterintuitive phenomenon is because due the stochastic driving does not act on the external fields in our model, but on the strength of the interaction, which favors the ordered phase with spontaneous symmetry breaking. All cases with different $\mathcal{D}$, $M(t)$ always ends up with an algebraic decays $\sim t^{-\chi}$, whose exponent $\chi$ barely depends on $\mathcal{D}$ expect for the case with small $\mathcal{D}$ where the large statistical error makes it difficult to determine the accurate value of $\chi$ (see inset of Fig.\ref{fig:fig2} a).

Due to the absence of dissipation in our model, one may wonder whether the long-time behavior depends on the initial state despite the energy nonconservation. To evaluate this, we fix $\mathcal{D}$  but choose different initial states as the ground states of the Hamiltonian.(\ref{eq:Hamspin}) with $\xi(t=0)=\mathcal{D}_0\neq \mathcal{D}$. As shown in Fig.\ref{fig:fig2} (b), after some relaxation time, the $M(t)$ starting from different initial states will converge into the same trajectory, which indicates that the initial state information has been washed out by the stochastic driving.  It is remarkable to notice that this conclusion holds even for a symmetry unbroken initial state with $m(t=0)=0$. As shown in Fig.\ref{fig:fig2} (b), $M(t)$ with $\mathcal{D}_0=0$ will also converge to the universal algebraic decay after an  extraordinarily long time, which suggests that even though $m(t)=0$ is a solution of the EOM.(\ref{eq:EOM}), it is unstable and nontrivial dynamics can be triggered by small temporal fluctuation. .

The spectrum function $h_k^z$ is crucial for the determination of the  undriven system properties. Until now, our focus was the dispersion spectrum of a 1D tight binding model with $h_k^z=-2J\cos k$. Here, we investigate additional spectrum functions, such as  a gapped spectrum as $h_k^z=\pm\sqrt{(2J\cos k)^2+\Delta^2}$, which can be realized by the imposition of a staggered chemical potential $\Delta\sum_i(-1)^i \hat{n}_i$ on the original Hamiltonian.(\ref{eq:Hamboson}).  Even though the equilibrium properties of  gapless and gapped systems are significantly different from each other, their long dynamics under stochastic driving seem qualitatively the same, as shown in Fig.\ref{fig:fig2} (c). More generally, one can select a  spectrum function where each $h_i^z$  is randomly sampled from a uniform random distribution with $h_i^z\in [-2J,2J]$. As shown in Fig.\ref{fig:fig2} (c), the algebraic decay also holds for such a random spectrum function.

{\it Finite-size effect --} The systems we considered so far are  sufficiently large ($L=5000$), a fact that allows us to neglect the finite-size effect within the time scale of our simulation. The dynamics of $M(t)$ in smaller systems have been shown in Fig.\ref{fig:fig2} (d), from which it is shown that the universal algebraic decay of $M(t)$ will not persist forever, instead, it will eventually approach a saturation value accompanied by small fluctuations. The saturation time linearly scales with the system size, suggesting that this is a finite-size effect. One can further investigate the dependence of the finite saturation value $\bar{M}(L)$ on the system size. As shown in the inset of Fig. 1 (e), $\bar{M}(L)\sim L^{-\beta}$, where $\beta=0.48(1)$ (close to the value $\frac 12$ !) is another ``critical'' exponent in our model.

{\it Stochastic driving protocols other than white noise:} Till now, we have only considered the situations where $\xi(t)$ fluctuates as a white noise. Therefore, it is essential to assess whether such a universal algebraic decay holds for other stochastic protocols. Consequently, we first consider a quasi-periodic driving protocol as a superposition of two periodic drivings with incommensurate periods, that shares some common features with the stochastic driving for $t\rightarrow \infty$.  For instance, we select $\xi(t)$ as:
\begin{equation}
\xi(t)=V_0[\cos(2\pi t+\varphi)+\cos(\sqrt{3} t)]
\end{equation}
where $V_0$ is the strength of the quasi-periodic driving, and $\varphi\in [0,2\pi]$ is a time-independent random phase. We define  $M(t)=\sqrt{\langle m^2(t) \rangle_{\bm\varphi}}$, where the ensemble average $\langle\cdot\rangle_{\bm\varphi}$ is performed over different $\varphi$. As shown in Fig.\ref{fig:fig2} (e), we can then find two distinguished dynamical behaviors: for small $V_0$ (e.g. $V_0=3J$), $M(t)$ exhibit a persistent oscillation with a constant amplitude, and for large $V_0$ (e.g. $V_0=7J$), the universal algebraic decay with $\chi=0.34(4)$ seems to reappear in the presence of such a quasi-periodic driving. This phenomenon reminds us of the Aubry-Andr\'e model, which is known to have a localization transition when increasing the strength of the incommensurate potential~\cite{Aubry1980}. It would be interesting to investigate whether similar transitions can occur in the time domain of our model with quasi-periodic driving.

Nonetheless, we now consider another stochastic driving protocol known as telegraph noise, where $\xi(t)$ randomly jump between two discrete values $-\mathcal{D}$ and $\mathcal{D}$ with a transition rate $\kappa$ (the transition probability per unit time), which measures the (inverse) correlation time of this colorful  noise ($\langle\xi(t)\xi(t')\rangle_{\bm{\xi}}\sim  e^{-\kappa|t'-t|}$)\cite{Gardiner2010}. As shown in Fig.\ref{fig:fig2} (f), $M(t)$  also exhibit an algebraic decay in the presence of the telegraph noise, with an exponent $\chi=\frac 13$ independent of the driving strength $\Delta$ and the transition rate $\kappa$, which indicates that these universal dynamics will still survive in the presence of colorful noise.

{\it Discussion --} Here, we propose a heuristic explanation of the universal exponent $\chi=\frac 13$ based on a hydrodynamics description for the coarse CDW order parameter $\phi(x,t)=\sum_{i\in \mathbb{X}}(-1)^i \langle \hat n_i \rangle$ where the summation is over sites within the fluid cell centered at x. The EOM of the hydrodynamics for $\phi(x,t)$ takes the general form of $\partial_t \phi=-D \nabla^2 \phi +\cdots$, where $D$ is the diffusion coefficient depending on the strength of the noise. If the higher order terms $\cdots$  are irrelevant, it describes a standard diffusion where each Fourier component $\phi_k(t)=\int dx e^{ikx}\phi(x,t)$ decays as $\phi_k(t)\sim e^{-Dk^2t}$, and the global CDW order parameter can be considered as a collective behavior of different k modes: $m(t)=\int dk \phi_k(t)$.   If the noise is  additive, thus $D$ is a time-independent constant, we have $m(t)\sim \int dk e^{-Dk^2t}\sim t^{-\frac 12}$, which agrees with the dynamics observed in systems driven by a stochastic external field\cite{Cai2017}. However, Eq.(\ref{eq:MFspin}) indicates that the strength of the effective noise $[m(t)\xi(t)]$ does not only depends on the bare stochastic driving, but also on the state of the system, and thus the diffusion coefficient is time-dependent $D(t)\sim m(t) \mathcal{D}$, and one can obtain:
\begin{small}
\begin{equation}
m(t)\sim \int dk e^{-\mathcal{D} m(t) k^2t}\sim [m(t)t]^{-\frac 12}
\end{equation}
\end{small}
which leads to  $m(t)\sim t^{-\frac 13}$ that agrees with our numerical observation.  Despite the simplicity of this argument, it doesn't explain why such dynamics is so universal. A systematic answer to this question  calls for an effective field theory description of our system in the Keldysh formalism\cite{Kamenev2011,Sieberer2016} that is augmented by a renormalization group analysis, which will be performed in  future studies.

It is interesting to compare  our results to other dynamical universality classes observed before.  In the KPZ model of growing interface in a strip geometry with width $L$, the roughness of the interface  $w$ initially increases algebraically with time as $w(t)\sim t^\chi$ ($\chi=\frac 13$)\cite{Huse1985,Huse1985b},  but eventually, due to the finite size effect, plateaus at some saturation value that scales with $L$ as $w(t\rightarrow\infty)\sim L^\beta$ ($\beta=\frac 12$)\cite{Kardar1986}. Despite the striking similarity of the exponents $\chi$ and $\beta$, a major obstacle for further comparisons between the KPZ and our model is a lack of characteristic length in our model due to its all-to-all coupling feature, and thus it is unclear how to characterize the dynamical critical exponent $z$ which measures the relationship between the characteristic length and time here.  Also, even though the KPZ universality class is also due to the conspiracy of stochasticity and nonlinearity, the noise in the KPZ equation is additive instead of multiplicative  as in our case.  Due to these important discrepancies, it is unclear whether these striking similarities  are just coincidence or whether there are deep reasons behind them. Other power-law decays with different powers have been discovered before in interacting quantum systems subjected to white noises\cite{Poletti2012,Cai2013,Ren2020}, where the interplay between the interaction, symmetry and noise plays an important role.  For instance, the algebraic decay observed in noisy quantum spin chain\cite{Cai2013} crucially depends on the presence of continuous symmetry (or equivalently the conservation law). Here, however, it is unclear whether the $\frac 13$ power-law universality class discussed above is related with some conservation quantities or continuous symmetries, since in our Hamiltonian.(\ref{eq:Hamspin}), the symmetry is  discrete ($Z_2$ type) instead of continuous.

Finally, we will add some remarks regarding the effect of spatial fluctuations, that, although are suppressed in our model, widely exist in local Hamiltonian with symmetry breaking phases. Even though an exact simulation of a non-equilibrium interacting quantum system is a formidable, if not impossible, task, one can estimate that in general, both the quantum and thermal fluctuations tend to thermalize the system within a typical time scale $t_\Delta$. If $t_\Delta$ is much longer than the typical time scale of our Hamiltonian dynamics $t_s\sim \mathcal{O}(J^{-1})$, then it is natural to expect that the universal dynamics discussed above can still be observed in the prethermal regime.  In addition, motivated by the equilibrium phase transition theory where the mean-field method works for systems with dimension above four, we speculate that the $\frac 13$ power-law universality class observed in our infinite dimensional model, holds for systems with sufficiently high dimension. However, an estimation of the upper critical dimension $D_c$ below which the dynamics is qualitatively different from the $\frac 13$ power-law is a highly non-trivial problem, which calls for a renormalization group analysis in the framework of non-equilibrium field theory. The upper critical dimension in our model, if exists, does not have to be four, since our model is far from equilibrium.

{\it Conclusion and outlook  --} In conclusion, our findings show that despite the simplicity of the mean-field EOM.(\ref{eq:EOM}), it can manifest a remarkable dynamical universal class which is absent in equilibrium cases. Future developments will include an analytic explanation of the universality based on a non-equilibrium field theory and renormalization group analysis. Another important question pertains to  the generality of our conclusions: whether  dynamical behaviors other than the universal algebraic decays (for instance, exponential or stretched exponential decays\cite{Poletti2012}) exist in different parameter regimes of our model. If so, what is the critical behavior between these different dynamical phases?  Finally yet importantly, it is essential to understand the effect of spatial fluctuation on  universal dynamics beyond the mean-field treatment. Even though this question is extremely difficult to be answered in the context of quantum many-body systems,  numerical simulations on the classical systems may shed light on this problem\cite{Yao2020}.

{\it Acknowledgments}.---This work is supported by the National Key Research and Development Program of China (Grant No. 2020YFA0309000), NSFC of  China (Grant No.12174251),  Shanghai Municipal Science and Technology Major Project (Grant No.2019SHZDZX01).

%\bibliography{real}

\end{document}